\documentstyle[preprint,aps]{revtex}
\begin{document}
\draft
\title{Resonant Transport in Coupled Quantum Wells: a Probe for Scattering
Mechanisms}
\author{Y. Berk $^1$, A. Kamenev$^2$, A. Palevski$^1$,
L. N. Pfeiffer$^3$, and K. W. West$^3$ }

\address{$^1$School of Physics and Astronomy, Raymond and Beverly Sackler
Faculty of
Exact Sciences,\\
Tel Aviv University, Tel Aviv 69978, Israel\\
$^2$Department of Condensed Matter, The Weizmann Institute of Science,
Rehovot 76100, Israel.\\
$^3$ AT\&T Bell Laboratories, Murray Hill, New Jersey 07974}

\date{\today}
\maketitle

\begin{abstract}
We present a microscopical theory and experimental results concerning
resistance resonance in two tunneling coupled quantum wells with different
mobilities. The shape of the resonance appears to be sensitive to the small
angle
scattering  rate on remote impurities and to the electron--electron scattering
rate.
This allows the extraction of scattering parameters directly from the transport
measurements. The negative resonance in a Hall coefficient is predicted and
observed for the first time.
\end{abstract}


\narrowtext
The novel resistance resonance (RR) in two coupled quantum wells (QW) with
different mobilities was
recently suggested and discovered experimentally \cite{Palevski}.
The basic physical idea of this phenomenon is the following. One studies the
in--plane resistance of two QW's  as function
of a relative position of their energy levels (gate voltage),
by contacting to both of them.
If the  energy levels are far from each other, the tunneling is suppressed
and the resulting  resistance is given by,
${R}_{off} \sim (\tau^{tr}_1 +
\tau^{tr}_2)^{-1}$ (two resistors connected in parallel),
where $\tau^{tr}_i$'s  are the transport  mean free times in the corresponding
wells. The situation, however, is remarkably
different once the system is brought into the resonance
(energy levels coincide). In this case the wavefunctions form
symmetric and antisymmetric subbands, split by a tunneling gap.
As any electron is completely delocalized between two wells, the scattering
rate in each of these subbands is
$(\tau^{tr})^{-1} = (2\tau_1^{tr})^{-1} + (2\tau_2^{tr})^{-1}$.
The resulting resistance is given by ${R}_{res} \sim (2\tau^{tr})^{-1}$.
If the transport scattering rates of two QW's are different, one should find
the peak in the resistance whose relative amplitude is
\begin{equation}
{{R}_{res} - {R}_{off} \over {R}_{off}}
= {(\tau^{tr}_1 -\tau^{tr}_2)^2 \over 4 \tau^{tr}_1 \tau^{tr}_2}
\equiv A.
                                                             \label{A}
\end{equation}
This effect was indeed observed and reported
in a number of publications \cite{Palevski,Pal,Sakaki}.
In this letter we present a microscopical model of the RR, which includes
elastic scattering on a long--range remote impurity potential.
We also report the experimental measurements of the RR shape and
its temperature dependence, and we analyze them  within the presented
theoretical model. In addition, we have calculated and
measured the Hall coefficient in coupled QW's.
We show that at the resonant conditions the Hall coefficient exhibits a
local minimum, which may be well understood on a basis of a classical
two--band model.

The main messages which follow from the present investigation are the
following:
(i) at low temperature, the width of the RR is determined by a small angle
scattering time on remote impurities;
(ii) the temperature dependence of the RR indicates that
a shape of the resonance is  sensitive to the electron-electron
interactions, allowing determination of the electron--electron scattering
rate;
(iii) the resonance in a Hall coefficient is predicted and
demonstrated experimentally.

We first present a theoretical model of a transport in two coupled
QW's. In a basis of local states of each QW, the Hamiltonian
of the system may be written in the following matrix form:
$H=\hat a^{\dagger}_{\bf k}\hat H_{\bf k,p}\hat a_{\bf p}$, where
\begin{equation}
\hat H_{\bf k,p}=\delta_{\bf kp}  \left( \begin{array}{lr}
\epsilon_1+ ({\bf p}-e{\bf A})^2/(2m^*)   & \Delta/2 \\
\Delta^*/2  & \hskip -2cm      \epsilon_2+({\bf p}-e{\bf A})^2/(2m^*)
\end{array} \right) +
\left( \begin{array}{cc}
U_1({\bf p-k})   & 0  \\
0                &  U_2({\bf p-k})
\end{array} \right) \, ,
                                                                \label{ham1}
\end{equation}
\narrowtext
and   ${\bf k,p}$ are 2D  momentum of the electrons. In the last equation
$\epsilon_i(V_G)$ are bare quantized levels of
corresponding wells, which are  functions of a gate voltage, $V_G$.
The tunneling coupling (gap),
$\Delta$, is assumed to be in--plane momentum conserved and
energy independent.
A vector potential of the external field (electric and magnetic) is denoted by
${\bf A}={\bf A}({\bf r},t)$. Finally,
$U_i({\bf p-k})$ represents the elastic disorder in each layer. We shall assume
that impurity potentials in different wells are uncorrelated.
Inside each QW an impurity potential has a finite correlation length and
may be characterized by the two scattering times:
the full one (or small angle) and the transport one
\begin{equation}
\frac{1}{\tau_i}\propto \int |U_i({\bf p})|^2 d\Omega;\,\,\,\,\,
\frac{1}{\tau_i^{tr}}\propto \int |U_i({\bf p})|^2(1-\cos\theta) d\Omega,
\end{equation}
where  the integrations are carried out over the Fermi circles.

Now the model is specified completely and we  apply it first to the
calculation of a linear conductance. Using the Kubo formula, one has
\begin{equation}
\sigma= \int d\epsilon \frac{f(\epsilon)-f(\epsilon+\omega)}{2\pi S\omega}
\mbox{Tr} \langle \hat I_{\bf p} \hat G^+_{\bf p,k}(\epsilon+\omega)
\hat I_{\bf k} \hat G^-_{\bf k,p}(\epsilon) \rangle,
                                                            \label{kubo}
\end{equation}
where $\mbox{Tr}$ stays for both matrix and momentum indexes; $S$ is an area
of the structure.
A current operator, $\hat I_{\bf p}$, is $e{\bf p}/m$ times a unit matrix (if
all contacts are attached to the both wells).
Retarded and advanced Green functions of a system are defined as
\begin{equation}
\hat G^{\pm}_{\bf p,k}(\epsilon)=
<{\bf p}|(\epsilon-H\pm i \eta)^{-1}|{\bf k}>.
\end{equation}
Constructing the perturbation expansion over the impurity potential (the
second term in Eq.\ (\ref{ham1})), and solving the Dyson equation for an
average Green function, one obtains
to leading order in $(\epsilon_F\tau_i)^{-1}$, \cite{Abrikosov64}
$$
\langle \hat G^{\pm}_{\bf p,k}(\epsilon) \rangle =
\delta_{\bf kp}  \left( \begin{array}{lr}
\epsilon-\epsilon_1-\epsilon_{\bf p}\pm i/2\tau_1      & \Delta/2 \\
\Delta^*/2 & \hskip -1cm  \epsilon-\epsilon_2-\epsilon_{\bf p}\pm i/2\tau_2
\end{array} \right)^{-1}.
$$
Note that a tunneling coupling is taken into account in a non--perturbative
fashion, hence the final results should not be restricted by lowest orders in
$\Delta$. The conductivity, according to Eq.\ (\ref{kubo}) is given by a
diagram Fig.\ 1a, where the shaded triangle represents the renormalized
current vertex. To evaluate the latter
one should solve the matrix integral equation
schematically depicted on Fig.\ 1c,  \cite{Abrikosov64}. The calculation gives
the following result for the zero-frequency  resistance ($R=\sigma^{-1}$)
\begin{equation}
R= \frac{R_1 R_2}{R_1+R_2}  \left[ 1 +
A \frac{|\Delta|^2\tau^{tr}\tau^{-1}}
{(\epsilon_1-\epsilon_2)^2+|\Delta|^2\tau^{tr}\tau^{-1}+\tau^{-2}}
\right];
                                                               \label{res}
\end{equation}
$$\frac{1}{\tau}=\frac{1}{2\tau_1}+\frac{1}{2\tau_2};\hskip 2cm
\frac{1}{\tau^{tr}}=\frac{1}{2\tau_1^{tr}}+\frac{1}{2\tau_2^{tr}}, $$
where $R_i=(e^2 n_i\tau_i^{tr}/m)^{-1}$ are  resistances of each well and
the asymmetry coefficient, $A$, is defined by Eq.\ (\ref{A}).
The result, Eq.\ (\ref{res}), is valid if all relevant energies are much less
than the Fermi energy, $\epsilon_F$; this implies that the concentrations of
carriers in two QW's are close to each other, $|n_1-n_2|\ll n_i$.
For relatively clean case,
$|\Delta|^2\gg (\tau\tau^{tr})^{-1}$, Eq.\ (\ref{res}) confirms our
qualitative conclusions, drawn in the beginning. In the dirty case (the
opposite limit) the height of the resonance is suppressed.
Note, that the width of the resonance depends on the small angle
scattering time, $\tau$, although the resistances of each well are fully
determined by the transport times, $\tau_i^{tr}$. The physical nature of this
fact is the following. Any elastic scattering process (including the small
angle scattering) leads to a mixing between the states of symmetric and
antisymmetric subbands (according to classification in clean wells). Not too
far from the exact resonance (say $\epsilon_1-\epsilon_2\approx\Delta$) the
wavefunctions of clean wells are already mostly localized in one of the wells
(eg. ``symmetric'' in the upper one and ``antisymmetric'' in the lower one).
In this case they are sensitive only to scatterers in the
corresponding well and the resonance is destroyed. The above mentioned mixing
changes the situation, making the exact
eigenfunction of dirty wells delocalized. As a result the
resonance appears to be broader, than in the case without small angle
scattering.
The relative amplitude is  determined only by transport quantities and is not
affected by the latest.

The Hall coefficient is given by the two diagrams, one of which is depicted
in a Fig. 1b,  \cite{Altshuler85}. We present here only the
result for the short range impurity potential ($\tau_i=\tau_i^{tr}$)
\begin{equation}
R_H= \frac{R_{H,2} R_1^2+ R_{H,1} R_2^2}{(R_1+R_2)^2}  \left[ 1 -
A\frac{2\tau_1\tau_2}{\tau_1^2+\tau_2^2}
|\Delta|^2
\frac{(\epsilon_1-\epsilon_2)^2+|\Delta|^2+3\tau^{-2}}
{\left[ (\epsilon_1-\epsilon_2)^2+|\Delta|^2+\tau^{-2} \right]^2}
\right],
                                                               \label{hall}
\end{equation}
where $R_{H,i}=(n_i e)^{-1}$ is a Hall coefficient of each QW.
We shall discuss the physics of the last expression later,
when presenting the experimental results.

The double QW structure was grown on N$^+$ GaAs substrate
by molecular-beam epitaxy and
consisted of two GaAs wells 139 $\AA$ width separated by a 40 $\AA$
Al$_{0.3}$Ga$_{0.7}$As barrier.
The electrons were provided by remote delta-doped donor layers set back
by 250 $\AA$ and 450 $\AA$ spacer layers from the top and the bottom
well correspondingly.
In order to obtain the difference in the mobilities, an enhanced amount of
impurities  was introduced at the upper edge of the top well (Si, 10$^{10}$
cm$^{-2}$).
Measurements were done on
10$\mu$m-wide and 200 $\mu$m-long channels with Au/Ge/Ni Ohmic
contacts.  Top and  bottom gates
were patterned using the standard photolithography fabrication
method.
The top Schottky gate covered 150 $\mu$m of the channel.
The data were taken using a lock-in four terminal
techniques at $f$= 11 Hz. The voltage probes connected to the
gated segment of the channel  were separated by 100 $\mu$m.

The application of the upper
gate voltage allows us to sweep the potential profile
of the QW's through the resonant configuration.
The variation of the resistance vs. upper gate voltage
is plotted in Fig. 2 (circles). The data were obtained
at the T=4.2 K for the bottom gate voltage  $V_{GB} = 0.5 V$.
The resistance resonance is clearly observed at $V_G\approx -0.6 V$.

In order to compare the experimental data with the theoretical
formula, Eq.\  (\ref{res}), one has to establish the correspondence between the
gate voltages and the energy levels, $\epsilon_i$. The latter was found,
using the known density of states and dc electrical capacitances between the
QW's and corresponding gate electrodes. The experimental values of these
capacitances were
established using the Hall measurements in the regime of the complete
depletion of the top QW, and are given by
$ C_1 = 4.53 \times 10^{-8} F cm^{-2}$
for the upper gate and $ C_2 = 1.79 \times 10^{-8} F cm^{-2}$ for the bottom
gate, which are extremely close to the theoretical estimates.
The complementary measurements of the resistance and Hall coefficient far
from the resonance allow us to determine the following parameters of our
structure (as grown, i.e., $V_G = V_{GB}=0$ and $T=4.2K$):
$\mu _1 = 47,000 cm^2/Vsec$ \cite{foot1}, $\mu_2 = 390,000 cm^2/Vsec$,
$n_1 = 4.7 \times 10^{11} cm^{-2}$, $n_2 = 2.5 \times 10^{11} cm^{-2}$.
The quantum mechanical calculation of the tunneling gap results in
$\Delta=0.55$meV; a very similar value for an identical structure was found
experimentally \cite{boeb}. The single fitting parameter, which was not
determined by independent measurement is a small angle scattering rate,
$\tau^{-1}$. The best fit (solid line in Fig. 2) was achieved for
$\tau^{-1} = 1.3 meV$. This value implies the ratio between transport and
small angle scattering times to be equal to 4.7, which is in a very good
agreement with the measurements, using Shubnikov--de Haas oscillations,
\cite{Cole}. To our knowledge this is the first time, when the small angle
scattering rate was determined in a pure (zero magnetic field)
transport experiment.

The same fitting procedure was applied to a set of the resistance resonance
data  within the temperature range 4.2 -- 60K, see Fig. 3.
Amongst the independently measured parameters, only the mobility of a clean QW,
$\mu_2$, exhibits pronounced temperature dependence, which is consistent
with previously reported experimental data \cite {Pfeiffer}.
The temperature dependence of the fitting parameter,
$\tau^{-1}(T)$, is plotted by circles in an inset to Fig. 3.
At low temperature it may be well approximated by
the following relation (the solid line in the inset):
\begin{equation}
\tau^{-1}(T)=\tau^{-1}+3.0\, T^2/\epsilon_F,
\end{equation}
where $\epsilon_F=10.9$ meV is the Fermi energy and $\tau^{-1}=1.3$ meV
is a zero temperature scattering rate,
associated with a small angle scattering on the remote impurities.
We tend to attribute the quadratic  dependence of the scattering rate on
temperature   to an electron--electron (e--e)
interactions. Indeed, in a clean limit ($\tau^{-1}\ll T \ll \epsilon_F$),
the e--e scattering rate is given by \cite{Ashcroft76}
$\tau^{-1}_{ee}=\alpha T^2/\epsilon_F$,
where dimensionless coefficient $\alpha$ is of order of unity.
The e--e interactions  do not change the resistances of each well separately,
due to conservation of the total momentum of an electronic system.
Therefore, the interactions  do not influence the resistance at the resonance
and very far from it.
In the intermediate region, however, e--e interactions cause mixing between
symmetric and antisymmetric subbands, making the resonance broader. In this
sense it plays a role very similar to that of a small angle scattering (see
discussion after Eq. (\ref{res})). Following this argument, we assume
that the e--e scattering rate, enters the expression in the same way as a
small angle one. The last suggestion
requires some additional theoretical treatment; however, if verified, it
provides a powerful method of measuring of e--e scattering rate.

The Hall effect measurements are necessary to establish parameters of the
structure. They are, however, interesting due to a presence of
a "negative" resonance in a Hall coefficient. The experimental data for
the Hall coefficient, $R_H$, at $T=4.2$K and magnetic field less than $0.05$ T
(the region,  where a Hall voltage is linear with field)
is presented in Fig.\, 4 (circles).
The theoretical curve (see Eq. (\ref{hall})) is also plotted on the same graph
by a solid line.
The nature of the "negative" resonance may be easily understood
using a classical two--band model \cite{Ashcroft76}. According to this model,
two  bands having concentration of carriers $n_i$ and transport times
$\tau_i^{tr}$,
exhibit the  following Hall coefficient
\begin{equation}
R_H=\frac{1}{e} \frac{n_{1}(\tau^{tr}_1)^2+ n_{2}(\tau^{tr}_2)^2}
{(n_1\tau^{tr}_1+n_2\tau^{tr}_2)^2}.
\end{equation}
Far from the resonance the role of two bands are played by two QW's, thus in
this case $n_i$ and  $\tau^{tr}_i$ are characteristics of  uncoupled wells
(cf. Eq. (\ref{hall})).
In the exact resonance the two bands are symmetric and
antisymmetric subbands, which obviously have the same transport times,
$\tau^{tr}$,
and practically  the same concentrations, $n$
($\Delta\ll\epsilon_F$); thus in the resonance, $R_H=(2en)^{-1}$ (in agreement
with Eq. (\ref{hall})).
If the concentrations in the two wells differ from each other not too much
($n_1\approx n_2\approx n$), the
resonance value of the Hall coefficient is strictly less than the
off-resonance one.
Another prediction of the simple two band model is the dependence of a Hall
coefficient on a magnetic field \cite{Ashcroft76}. This was also observed
experimentally in a full agreement with a model, confirming that a classical
two--band model is applicable to our structure.

We have benefited from the useful discussions with A. Aronov,
O. Entin, V. Fleurov, Y. Gefen and Y. Levinson.
The experimental research was supported by Israel Academy of Sciences
and Humanities.
A.K. was supported by the
German--Israel Foundation (GIF) and the U.S.--Israel Binational
Science Foundation (BSF).

\figure{Fig. 1. Diagrams for a conductivity (a) and a Hall coefficient (b),
current vertex renormalization due to a small angle scattering (c).
Full circle -- bare current vertex; dashed line -- impurity scattering.}
\label{ff1}

\figure{Fig. 2. Resistance Resonance (RR) curves: circles -- experimental data,
solid line -- theoretical calculation.}

\figure{Fig. 3. The set of RR curves at different temperatures. The inset
shows the variation of $\tau^{-1}(T) - \tau^{-1}(0)$  vs. temperature.
The circles denote the values deduced from analysis
of experimental data,
the solid line represents  $3.0\, T^2/\epsilon_F$ [meV].}
\figure{Fig. 4. Hall coefficient vs. gate voltage: circles --
experimental data, solid line -- theoretical calculation.}

\end{document}